\documentclass[twocolumn]{aastex63}
\usepackage{color,comment,multirow,booktabs,upgreek,amsmath}
\graphicspath{{./}{figures/}}

\newcommand{\cmm}{\ifmmode{\rm cm^{-2}}\else{$\rm cm^{-2}$}\fi}
\newcommand{\hi}{\ifmmode{\rm HI}\else{H\/{\sc i}}\fi} 
\newcommand{\glon}{\ifmmode{\ell}\else{$\ell$}\fi} 
\newcommand{\glat}{\ifmmode{b}\else{$b$}\fi}
\newcommand{\vlsr}{\ifmmode{V_\mathrm{LSR}}\else{$V_\mathrm{LSR}$}\fi}
\newcommand{\vwind}{\ifmmode{V_\mathrm{w}}\else{$V_\mathrm{w}$}\fi} 
\newcommand{\dg}{\ifmmode{^\circ}\else{$^\circ$}\fi} 
\newcommand {\kms}{\ifmmode{\rm km \, s^{-1}}\else{$\rm km \, s^{-1}$}\fi} 
\newcommand {\mo}{{\rm M}_\odot}

\newcommand{\vout}{\ifmmode{V_\mathrm{out}}\else{$V_\mathrm{out}$}\fi}

\defcitealias{McClure-Griffiths+13}{McG13}
\defcitealias{DiTeodoro+18}{DiT18}

\begin{document}

\title{Observation of  Acceleration of \hi\ Clouds Within the Fermi  Bubbles}%

\correspondingauthor{Felix.~J. Lockman}
\email{jlockman@nrao.edu}

\author[0000-0002-6050-2008]{Felix J.\ Lockman}
\affiliation{Green Bank Observatory, Green Bank, WV 24944, USA}

\author[0000-0003-4019-0673]{Enrico M. Di Teodoro}
\affiliation{Research School of Astronomy and Astrophysics - The Australian National University, Canberra, ACT, 2611, Australia}

\author[0000-0003-2730-957X]{N. M. McClure-Griffiths}
\affiliation{Research School of Astronomy and Astrophysics - The Australian National University, Canberra, ACT, 2611, Australia}

\begin{abstract}

The $\sim200$ \hi\ clouds observed to be entrained in the Fermi Bubble wind show a trend of increasing maximum $|\vlsr|$ with Galactic latitude.  We analyze previous observations and present new data from the Green Bank Telescope that rule out systematic effects  as the source of this phenomenon.  Instead, it is likely evidence for acceleration of the clouds.  The data suggest that clouds 
in the lower 2 kpc of the Fermi Bubbles, within the Bubble boundaries established from X-ray studies, have 
an outflow velocity that rises from   $\approx 150 - 200$ \kms\ close to the Galactic Center and reaches 
$\approx 330$ \kms\  at a distance of $2.5 - 3.5$ kpc. 
  These parameters are also consistent with the kinematics of UV absorption lines from highly ionized species observed against two  targets behind the Fermi Bubbles at $b = -6\fdg6$ and $b = +11\fdg2$.  The implied neutral cloud lifetime is 4 -- 10 Myr.

\end{abstract}

\keywords{Galaxy: center -- Galaxy: nucleus -- ISM: clouds -- ISM: jets and outflows -- ISM: kinematics and dynamics}

\section{Introduction}

Energetic processes in the center of the Milky Way, arising from either star formation or active galactic nuclei (AGN) activity, have created two lobes extending to latitude $|b|\approx 55\dg$ above and below the Galactic Center.  Although evidence for a nuclear wind was being discussed many years before the launch of the Fermi $\gamma-$ray telescope \citep[e.g.,][]{Bland-Hawthorn_Cohen03, Veilleux+05,Keeney+06} { the structures} are now generally referred to as the Fermi Bubbles, clearly detected in  $\gamma-$ray emission \citep{Su+10}.  
Nuclear winds are ubiquitous in star-forming galaxies \citep{Veilleux+05}, and the Fermi bubbles give us the opportunity to study this phenomenon at a level of detail  not possible for more distant systems.
Associated components are visible in broad-band emission across the electromagnetic spectrum \citep{Dobler&Finkbeiner08,Kataoka+13,Carretti+13}.  
The Bubbles are also coincident with a void in the extended Galactic neutral hydrogen (\hi) layer at Galactocentric distances $R\lesssim 2.4$ kpc \citep{Lockman84,Lockman_McClureGriffiths16}.   

Spectroscopic observations of 21cm \hi\ emission  have detected material entrained in the bubbles at radial velocities exceeding 300 \kms \citep{McClure-Griffiths+13,DiTeodoro+18}  and measurements of ultra-violet (UV) absorption lines along paths through the bubbles reveal hot gas at similar large non-circular velocities in many stages of ionization  \citep{Fox+15,Savage+17}.  
Within { $|b| \lesssim 11\dg$} of the Galactic plane the 21cm and UV spectra have similar kinematics and have been modeled by a symmetric biconical outflow with constant velocity $\vout\simeq330 \, \kms$ through an opening angle of $140\dg$  centered on the Galactic Center \citep{DiTeodoro+18}.  
In this paper we show evidence from the 21cm \hi\  surveys that neutral clouds entrained in the Fermi Bubble wind appear to be accelerating from velocities near 150 \kms\ close to the Galactic Center to 330 \kms\ at a distance of several kpc.

The papers is organized as follows. 
In \autoref{sec:HI_Fermi} we describe 21cm \hi\ and UV spectra of gas entrained in the Fermi Bubble wind and present new observations made with the Green Bank Telescope (GBT).  
These data reveal a kinematic anomaly in the form of an absence of gas at $|\vlsr| > 200$ \kms\ within 4\dg\ of the Galactic plane.  
\autoref{sec:constant_vel} compares the observed distribution of velocities with that expected from a constant-velocity nuclear wind and concludes that there is a discrepancy.  \autoref{sec:systematics} considers a set of possible systematic effects that might produce the kinematic anomaly and concludes that the anomaly is real.  
\autoref{sec:acceleration} presents a simulation that matches the observed kinematics when  clouds are constantly accelerated for the first few kpc away from the Galactic Center. 
The paper concludes with a discussion in \autoref{sec:discussion}.

\begin{figure}
    \centering
    \includegraphics[width=0.5\textwidth]{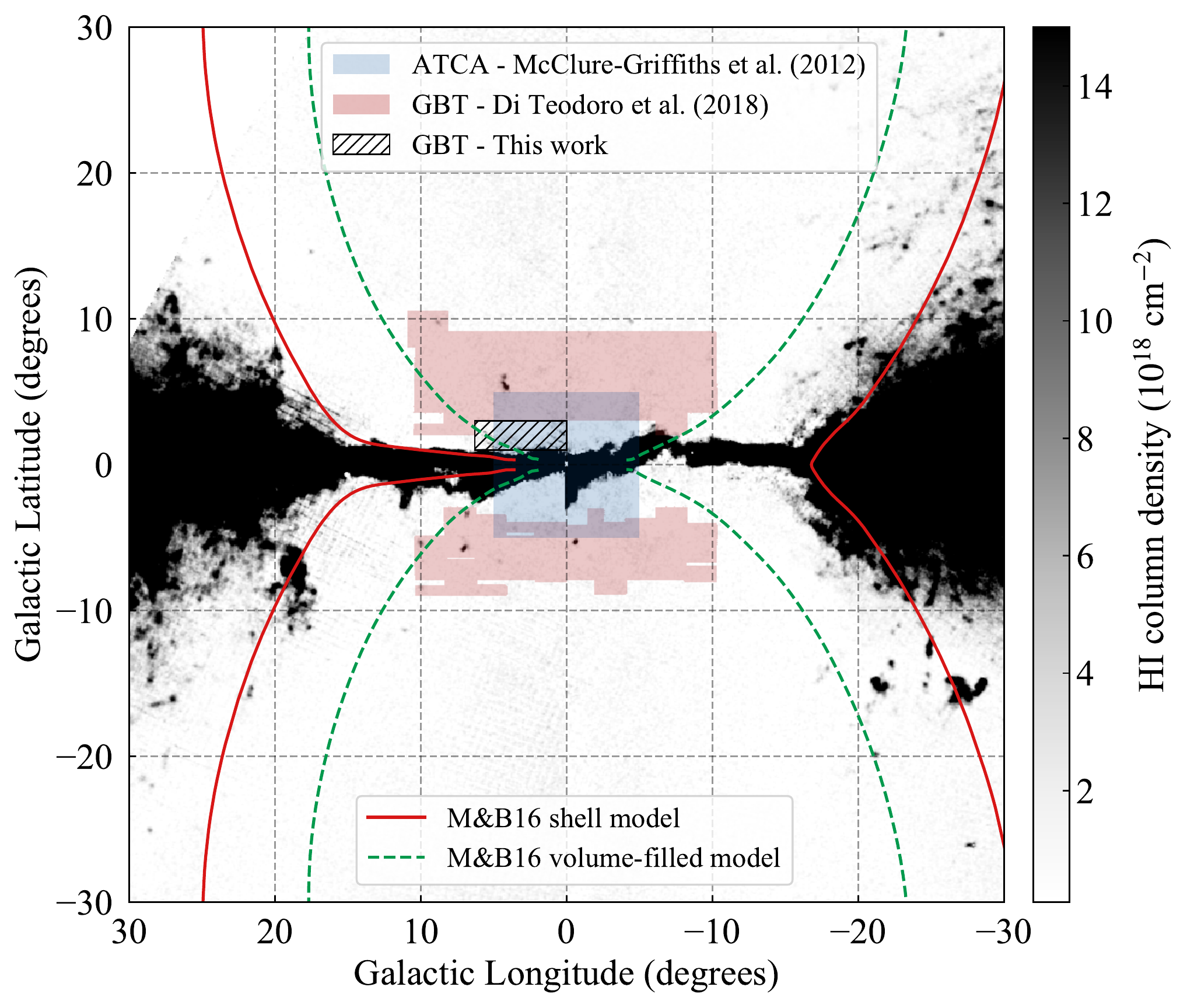}
    \caption{The greyscale is proportional to the \hi\ column density at the Galactic tangent points where ${\rm R = R_0 \ sin(\ell)}$ \citep{Lockman_McClureGriffiths16}.  
    The red solid curves and the green dashed curves mark the shell and volume-filled regions of the Fermi bubble from the model of \citet{MillerBregman16} derived from X-ray emission.  Hot, shocked wind material lies within the volume marked by the dashed green curves, while the outer shell includes shocked interstellar and circumgalactic material.  The area studied in the ATCA survey \citep{McClure-Griffiths+13} is marked by a rectangle $10\dg \times 10\dg$ around the Galactic Center, while the other shaded areas above and below  mark regions covered in the GBT survey \citepalias{DiTeodoro+18}, and the new observations presented here (crosshatched region).}
    \label{fig:HI_image}
\end{figure}

%----------------------------------------------------------
\section{Gas Entrained in the Fermi Bubbles}
\label{sec:HI_Fermi}

\autoref{fig:HI_image} shows the \hi\ emission along the tangent points of the Milky Way from \citet{Lockman_McClureGriffiths16} and superimposed on it the 
shell (red line) and volume-filled (green-dashed line) models
for the Fermi Bubble X-ray emission \citep{MillerBregman16}.  
The $10\dg \times 10\dg$ box centered on the Galactic Center (pink shaded) marks the region surveyed in \hi\ by the Australia Telescope Compact Array (ATCA) which first detected 86 \hi\ clouds entrained in the Fermi Bubble wind \citep[][hereafter \citetalias{McClure-Griffiths+13}]{McClure-Griffiths+13}.  
The red-shaded larger irregular boxes farther from the plane mark the regions surveyed by the 
Green Bank Telescope (GBT) which detected an additional 106 clouds \citep[][hereafter \citetalias{DiTeodoro+18}]{DiTeodoro+18}. 
\autoref{tab:surveys} compares the properties of those two radio \hi\ surveys (columns 2 and 4).  
Note that column density $N_\mathrm{\hi}$ used here always refers to the average over the resolution element of the observation.

\begin{deluxetable*}{lcccc}
\tablecaption{Properties of Galactic Center \hi\ surveys used in this work. 
\label{tab:surveys}}
\tablehead{
\colhead{Property}  & 
\colhead{ATCA$^1$}  & 
\colhead{\hspace{10pt}GASS$^2$}\hspace{10pt} &
\colhead{\hspace{10pt}GBT$^3$}\hspace{10pt} & 
\colhead{\hspace{10pt}new GBT$^4$}\hspace{10pt}}
\startdata
Angular Resolution 	  & $2\farcm4$ 	&	16\farcm2	& $9\farcm5$ & 9\farcm1 \\
Longitude range       & $\pm 5\dg$  & $\pm 12\arcdeg$ & $\pm 10\dg$ & $0\arcdeg - 6\arcdeg $\\
Latitude range        & $\pm5\dg$   & $\pm12\arcdeg$ & -8\dg $\lesssim b \lesssim -4\dg$ & $1\arcdeg - 3\arcdeg $ \\
		                            &               &                 & $+3\dg \lesssim b \lesssim +9\dg$ \\
Velocity range LSR (\kms) & $-309<\vlsr<+349$  & $\pm470$ & $\pm 670$  & $\pm656$  \\
Noise rms $\sigma_{T_b}$ (mK)\tablenotemark{a} & 700  & 57 & 30\tablenotemark{b} & 35 \\
$N_\mathrm{\hi}$ lower limit ($\cmm$)\tablenotemark{c}  &  $2.1 \times 10^{19}  $ & $1.6 \times 10^{18} $  & $9.0 \times 10^{17}$ & $1.0 \times 10^{18} $ \\ 
\noalign{\vspace{5pt}}
\enddata
\vspace{5pt}
$^1$\citet{McClure-Griffiths+13}; $^2$\citet{McClure-Griffiths+09}; $^3$\citet{DiTeodoro+18}; $^4$This work.
\tablenotetext{a}{In a 1 \kms\ channel.}
\tablenotetext{b}{The range is 23 mK to 40 mK over the mapped area.}
\tablenotetext{c}{Averaged over a resolution element for a Gaussian line with amplitude $A=3\sigma_{T_b}$ and FWHM = 30 $\kms$.}
\end{deluxetable*}

When all observational effects are taken into account, the neutral cloud population entrained in the Fermi Bubbles is nearly symmetric with respect to Galactic latitude, Galactic longitude, and \vlsr, and can be modelled as arising from a  filled bi-symmetric cone with only two free parameters: a maximum opening angle $\alpha$,  and a constant radial outflow velocity \vout\ \citepalias[see][for details]{DiTeodoro+18}.  
The outflowing gas clouds must be distributed throughout the cones,  because the \hi\ clouds in  any area (and the ionized components observed in UV absorption) span a range of velocity implying a range of ejection angle $\phi$,  where  $\phi$ is measured from the plane in the vertical direction.
The gas we observe is presumably entrained in a much faster hot wind whose kinematic structure is not observable.

The \hi\ clouds are detected down to a $|\vlsr|$ limited by confusion with unrelated lower-velocity \hi\ emission.  
In practice, the lower limit on $|\vlsr|$ of detectable \hi\ clouds is set by the sensitivity of the data and the algorithms used to identify and measure the clouds, which differ between \citetalias{McClure-Griffiths+13} and \citetalias{DiTeodoro+18}.  Generally speaking,  measurable cloud velocities are limited to $|\vlsr| \gtrsim  50$ \kms\ for the ATCA data and $\gtrsim75$ \kms\ for the GBT.  
This issue of confusion will be discussed further in \autoref{sec:confusion}.

\begin{figure*}
    \centering
    \includegraphics[width=0.48\textwidth]{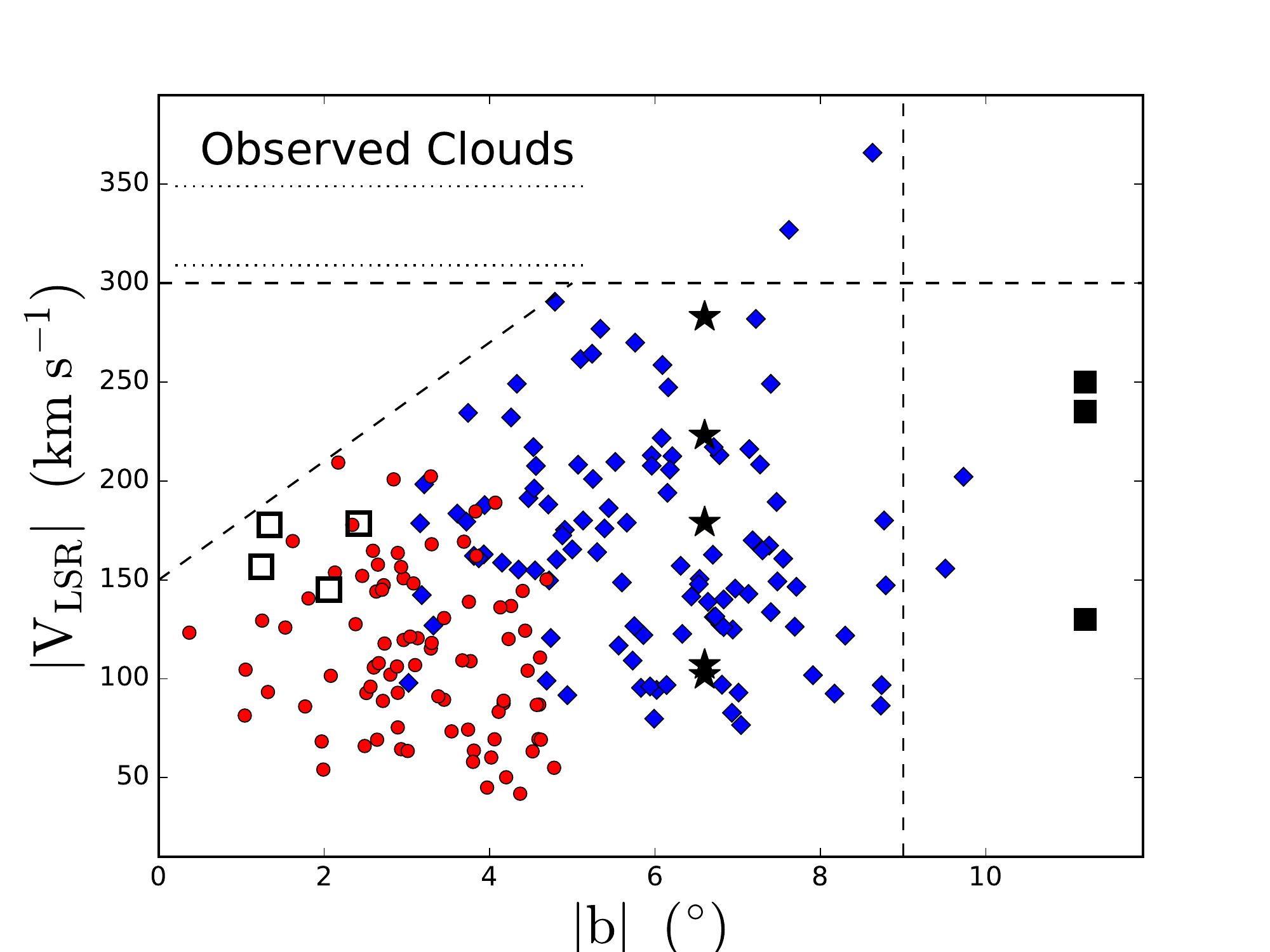}
    \hspace{0.02\textwidth}
    \includegraphics[width=0.48\textwidth]{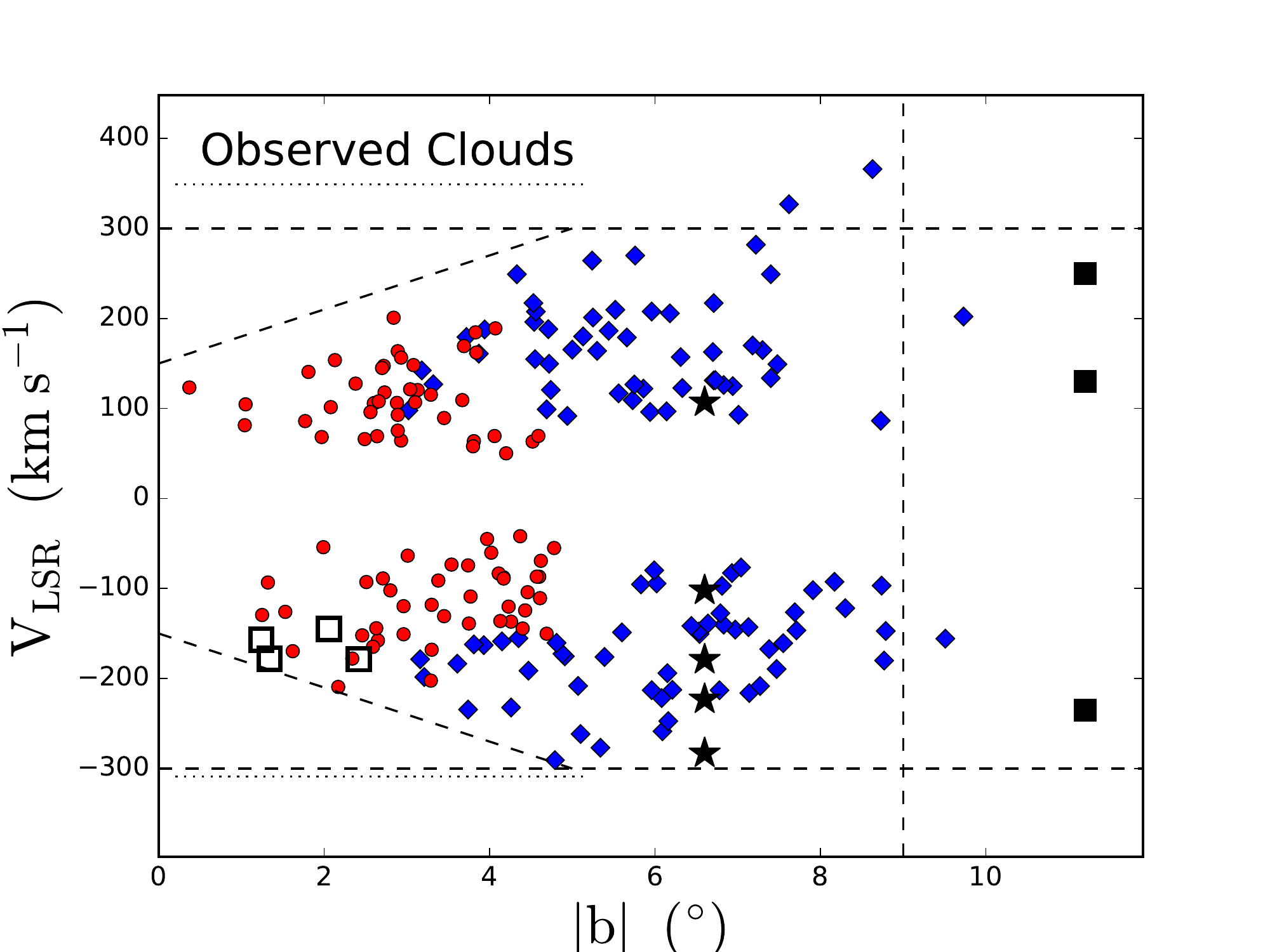}
    \caption{\emph{Left panel:} the $|\vlsr|$ of clouds detected in the ATCA survey (\citetalias{McClure-Griffiths+13}; red circles) and the GBT survey (\citetalias{DiTeodoro+18}; blue diamonds) as a function of the absolute value of the Galactic latitude, $|b|$.  Dotted  lines show the negative and positive $|\vlsr|$ limits, respectively,  of the ATCA survey over its latitude range. 
    The GBT survey is complete to  $\pm 670$ \kms.  Stars show velocities of five high-velocity  UV absorption line components detected towards a distant star behind the Fermi Bubble at $(\ell,b) = (1\fdg67,-6\fdg63$) \citep{Savage+17}, while filled  rectangles show the three high-velocity UV absorption components detected toward an AGN at $(\ell,b) = (10\fdg4,+11\fdg2$) from \citet{Fox+15}.  Open rectangles mark clouds discovered in the new GBT data described in this paper. Fiducial dashed lines at 300 \kms, and from 150 \kms\ at $|b| = 0\dg$ to 300 \kms at $|b| = 5\dg$, outline the region where clouds should be detected from a constant V$_\mathrm{out} = 330$ \kms\ wind, but are not.  This defines  the  kinematic anomaly.   At positive latitudes the GBT survey is complete at $b \leq +9\dg$ (dashed vertical line).  At negative latitudes the survey is complete to  only $b \geq -8\dg$. 
    \emph{Right panel:} Same as left panel, but for positive and negative velocities separately, and a line from $-150$ \kms\ to $-300$ \kms. The anomaly appears in both positive and negative velocity clouds.}
    %\edt{I have put together these two plots, as they basically show the same thing.}}
    \label{fig:AbsV_absb}
\end{figure*}

\citetalias{McClure-Griffiths+13} fitted a kinematic model to the ATCA data, and \citetalias{DiTeodoro+18} modelled the GBT data. 
Both the ATCA and GBT cloud samples are consistent with a cone opening angle $\alpha \approx 140\dg$, but analysis of the lower $|b|$ ATCA survey found $\vout = 220$ \kms\ while the GBT data required a model with a significantly higher $\vout = 330$ \kms. 

The reason for the difference is clear from \autoref{fig:AbsV_absb}. The left panel shows the $|\vlsr|$ for the combined ATCA and GBT samples plotted against the absolute value of the Galactic latitude, $|b|$.  Dotted and dashed lines show the positive and negative \vlsr\ limits of the ATCA survey over its latitude range.  The GBT survey covered $\pm670$ \kms.   It is evident that the maximum $|\vlsr|$ rises from about 125 \kms\ at $|b| \sim 1\dg$, the smallest latitudes where discrete clouds have been detected, to $\sim 300$ \kms\ around $|b| = 5\dg$.  
The ATCA survey over $|b| \leq 5\dg$  did  not detect any \hi\ clouds with ${\rm |V_{LSR}| > 210}$ \kms\ although this velocity range was included in its spectral coverage, while the higher-latitude GBT survey found clouds with  ${\rm |V_{LSR}| \gtrsim 330}$ \kms.  We refer to this difference as the kinematic anomaly.  
  
The right panel of \autoref{fig:AbsV_absb} shows the \hi\ clouds plotted separately for negative and positive \vlsr.  The absence of clouds at $|\vlsr| > 200$ \kms\ near the Galactic plane appears for both approaching and receding clouds.

\subsection{UV Absorption Spectroscopy}
\label{sec:UVabsorption}
Results from UV absorption line measurements on sightlines that pierce the Fermi Bubbles over these latitudes are included in \autoref{fig:AbsV_absb}.
The star LS 4825, a B1 Ib-II star located $21 \pm 5$ kpc from the Sun at ($\ell,b) = (1\fdg67$, $-6\fdg63$) shows discrete absorption features between velocities $-283$ \kms\ and +107 \kms\ over a range of species and  ionization states, including \ion{Si}{3}, \ion{C}{4}, 
\ion{Si}{4}, and \ion{N}{5}  \citep{Savage+17}.  
Above the plane, absorption spectra toward the AGN PDS 456 at ($\ell,b) = (10\fdg4, +11\fdg2$) shows UV metal-line absorption components 
(e.g., \ion{C}{2}, \ion{S}{2}, \ion{Si}{4})  from $-235$ \kms\ to +250 \kms\ \citep{Fox+15}.  
These absorption velocities are  consistent with the velocity range of the \hi\ clouds at a similar latitude.

\citet{Savage+17} note that the interstellar absorption toward a foreground star nearly aligned with LS 4825, but at a distance of only $ < 7 \pm 1.7$ kpc, is completely consistent with normal Galactic rotation, confirming that the high non-circular velocities arise in the Fermi Bubble and are not a general feature of the interstellar medium in the direction of the Galactic Center.

There are also measurements of high-velocity UV absorption lines through the Fermi Bubbles quite far from the Galactic Plane, at $28\fdg5 \leq |b| \leq 50\fdg3$ \citep{Bordoloi+17,Karim+18}, but these are well away from the areas we consider here.

\subsection{Additional 21cm HI Data}
\label{sec:new21cm}

There are significant differences between the \citetalias{McClure-Griffiths+13} ATCA and \citetalias{DiTeodoro+18} GBT surveys in sensitivity and angular resolution, as well as in the method of extracting clouds, which might be important when comparing  population properties.  
For this reason we did a search for \hi\ clouds in the GASS survey \citep{McClure-Griffiths+09, Kalberla+10} which covers the entire area of \autoref{fig:HI_image}, and made new observations with the GBT over 12 square degrees that overlap part of the ATCA survey but with greater sensitivity. 
Parameters of the GASS and new GBT observations are given in columns 3 and 5 of \autoref{tab:surveys}.  
The GBT observations and data reduction are described in \autoref{appendix:data}.

No clouds were detected in the GASS data that had not already been included in the ATCA and GBT samples.  
In contrast, while the new GBT observations recovered all clouds seen in the ATCA survey in the same part of the sky, four new clouds were detected as well.  
Their  properties are listed in \autoref{tab:newGBTdata}. 
The new GBT clouds have velocities in the range of those detected in the ATCA survey.  
These are shown as open rectangles in \autoref{fig:AbsV_absb} and do not resolve  the kinematic anomaly.
We conclude that the absence of \hi\ clouds at the highest $|\vlsr|$ at $|b| \lesssim 4\dg$ is not simply a result of different properties of the ATCA and GBT surveys.    

\section{Expectations from a constant velocity wind}
\label{sec:constant_vel}

\begin{figure}
	\centering
	\includegraphics[width=0.48\textwidth]{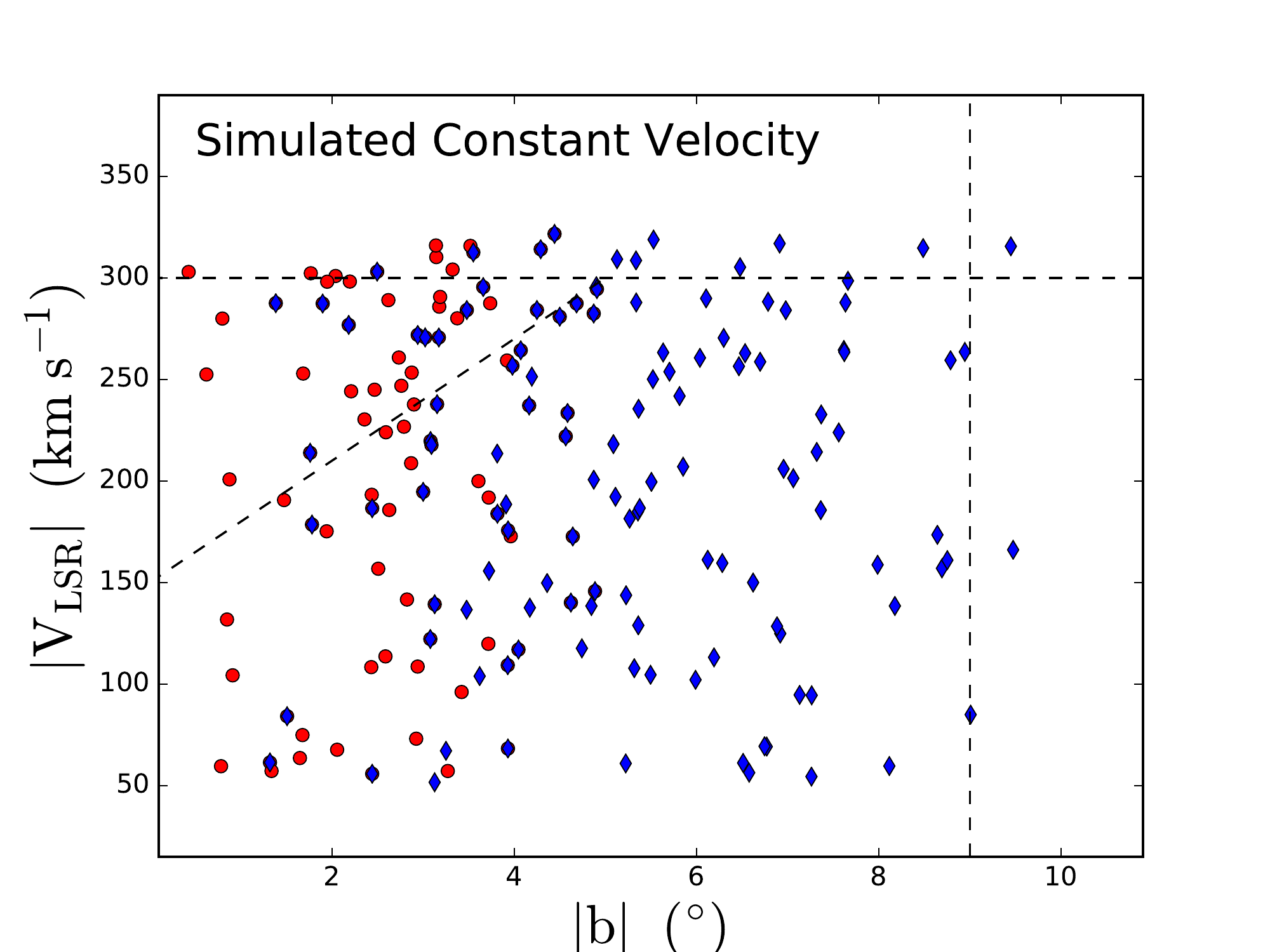}
	\caption{
		$|\vlsr|$ vs. $|b|$ for a simulated population of clouds filling two bicones with constant V$_\mathrm{out} = 330$ \kms\ and  a maximum opening angle $\alpha = 140\arcdeg$, the properties from \citetalias{DiTeodoro+18}.  Red circles show clouds that appear in the area covered only in the ATCA survey and blue diamonds show those that occur in regions covered by the GBT surveys. 
		The dashed fiducial lines are identical to those in Fig.~\ref{fig:AbsV_absb}.  
		This simulation does not match the observations as it produces many clouds with $|\vlsr| > 200$ \kms\ at low latitudes that are not observed.}
	\label{fig:Model_constVw}
\end{figure}

\autoref{fig:Model_constVw} shows the expected $|\vlsr|$ vs. $|b|$ for a populations of clouds drawn from a random sample originating in a bi-cone centered on the Galactic Center with an opening angle $\alpha = 140\dg$ and a constant radial outflow velocity $\vout = 330 \, \kms$, the parameters derived by \citetalias{DiTeodoro+18}, who analyzed only the GBT survey data.  The dashed fiducial lines are identical to those in \autoref{fig:AbsV_absb} and show that these parameters are not a good description of  the combined data set, as they predict a number of clouds with relatively high velocity at low latitude that are not present, especially in the ATCA data.

Before considering models incorporating acceleration, we examine the possibility that the kinematic anomaly arises from systematic effects.

\section{Possible Systematic Effects}
\label{sec:systematics}

\begin{figure*}
    \centering
    \includegraphics[width=0.45\textwidth]{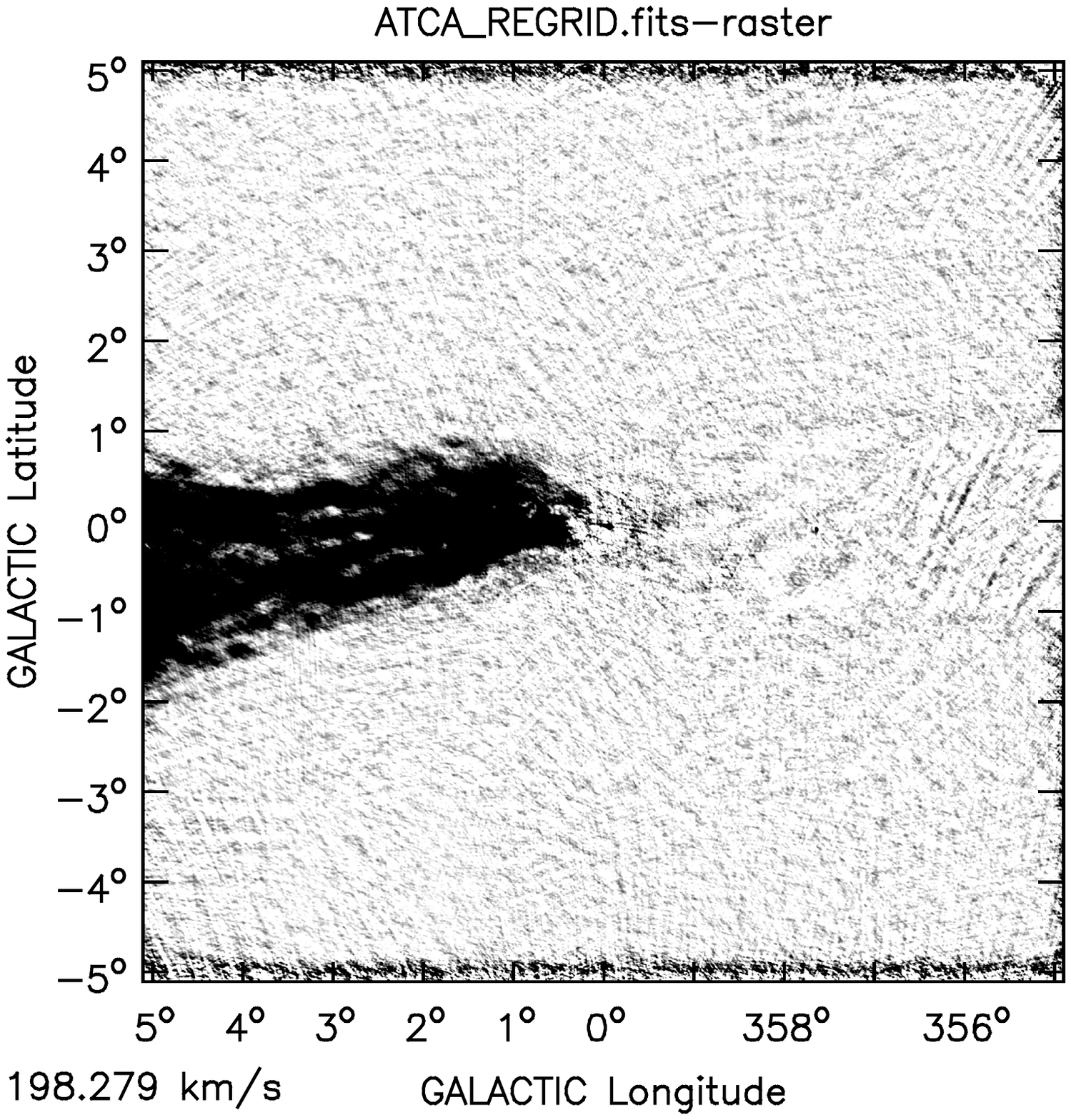}
    \hspace{0.04\textwidth}
    \includegraphics[width=0.45\textwidth]{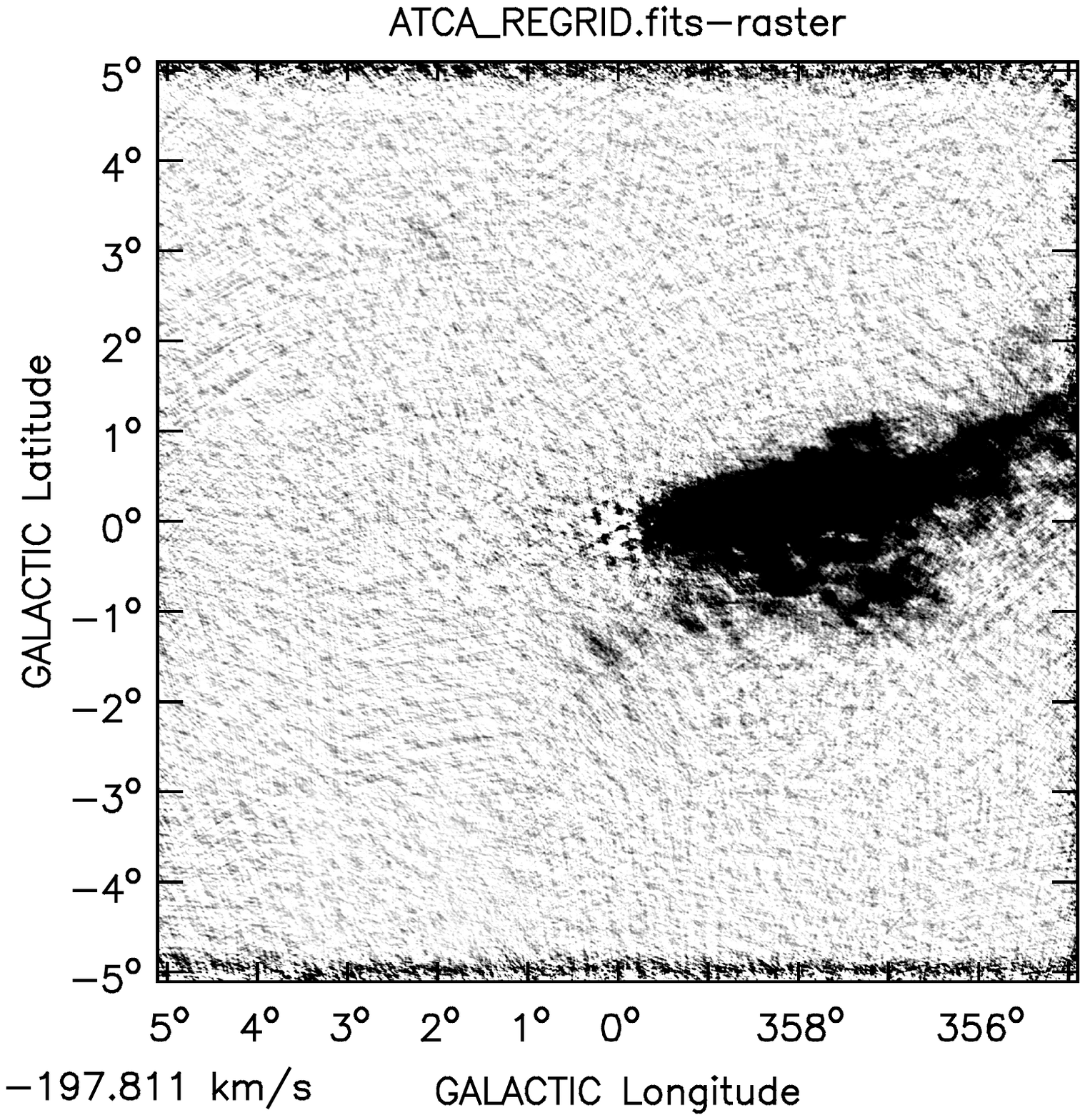}
    \caption{ATCA \hi\  survey data binned to 3 \kms\ at $\vlsr = +198$ \kms\ ({\it left panel}) and $\vlsr = -198$ \kms\ ({\it right panel}).  At these velocities one still sees emission from the tilted nuclear disk,  but it causes confusion over  $\lesssim 10\%$ of the area of the field.  At higher $|\vlsr|$ the confused region is correspondingly smaller.  There are no selection effects preventing detection of \hi\ clouds at $|\vlsr| \geq 200$ \kms\ over $\geq90 \% $ of the field.}
    \label{fig:ATCA_v_p_198}
\end{figure*}

%% New fits to lines 15 Sept 2019

\begin{deluxetable*}{lcccccccc}
\tablecolumns{9}
\tablecaption{Clouds detected in the new GBT Observations. Columns as follow: (1) Cloud assigned name; (2) Galactic longitude; (3) Galactic latitude; (4) LSR velocity; (5) Line width FWHM; (6) Maximum line brightness temperature; (7) Maximum \hi\ column density; (8) radius FWHM; (9) \hi\ mass. Columns (2) and (3) are from a Gaussian fit of the integrated total spectrum. Column (7) is from columns (5) and (6). Columns (8)-(9) are calculated assuming a distance from the Sun of 8.5 kpc. Uncertainties are propagated from the errors associated with the Gaussian fits. 
\label{tab:newGBTdata}}
\tablehead{\colhead{Name} & 
\colhead{$\ell$}  & 
\colhead{$b$} & 
\colhead{\vlsr} & 
\colhead{FWHM} & 
\colhead{$T_\mathrm{b, max}$} & 
\colhead{$N_\mathrm{\hi,max}$} & 
\colhead{$r$} & 
\colhead{$M_\hi$}\\
\colhead{} & \colhead{($\arcdeg$)} & \colhead{($\arcdeg$)} & \colhead{(\kms)} & \colhead{(\kms)} & \colhead{(K)} & \colhead{($10^{19}$ \cmm)} & \colhead{(pc)} & \colhead{($\mo$)} \\
\colhead{(1)} &  \colhead{(2)} & \colhead{(3)} & \colhead{(4)} & \colhead{(5)} & \colhead{(6)} & \colhead{(7)} & \colhead{(8)} & \colhead{(9)} 
}
\startdata
	\hline\hline\noalign{\vspace{5pt}}	
			\noalign{\smallskip}
G1.43+2.42-176 & 1.43 & 2.42 & $-176.4\pm0.8$  & $31.0\pm1.8$ & $0.18\pm0.01$ & $1.1\pm0.1$ & $19\pm4$ & $140\pm43$\\
G1.60+2.06-147 & 1.60 & 2.06 & $-147.2\pm0.7$  & $24.3\pm1.7$ & $0.28\pm0.02$ & $1.3\pm0.1$ & $23\pm2$ & $250\pm38$\\
G1.78+1.34-177 & 1.78 & 1.34 & $-177.4\pm0.4$  & $19.4\pm0.9$ & $0.37\pm0.02$ & $1.4\pm0.1$ & $26\pm2$ & $340\pm43$\\
G4.34+1.24-157 & 4.34 & 1.24 & $-157.4\pm0.5$ & $25.2\pm1.2$ & $0.24\pm0.01$ & $1.2\pm0.1$  & $32\pm3$ & $426\pm63$\\
\noalign{\vspace{5pt}}
\enddata
\end{deluxetable*}

In this section we consider a number of systematic observational effects that might be related to the absence of clouds with $|\vlsr| > 200 \ \kms$ at $|b| \lesssim 4\dg$, and conclude that none seems  likely to account for the kinematic anomaly. 

\subsection{Confusion}
\label{sec:confusion}

Clouds at low $|\vlsr|$ can be blended with unrelated foreground emission and become difficult to distinguish.  
The velocity where this confusion sets in varies somewhat with latitude but typically occurs at $|\vlsr|  \lesssim 50 \  \kms$.   
In addition to confusion from foreground \hi\ in the Galactic disk, there can be confusion from gas having non-circular motions associated with the inner Galaxy.  
The ATCA survey covered regions within a few degrees of the Galactic plane where the nuclear disk, 3-kpc arm, and other features with large non-circular velocities \citep[e.g.,][]{Oort77,Burton&Liszt78,Binney+91} make it impossible to  identify clouds associated with expansion of the Fermi Bubbles unambiguously.
But this effect is limited.

\autoref{fig:ATCA_v_p_198} shows channel maps for ATCA data at $\vlsr = \pm198$ \kms.  
Although within a few degrees of the Galactic plane there is significant emission at some longitudes,  $\approx 90\%$ of the field has no confusing emission preventing the detection of \hi\ clouds at the given velocity.  
At ever larger $|\vlsr|$ the confusing emission occupies an ever smaller area, reaching $\approx 4\%$ at  $|\vlsr| = 250$ \kms\ and becoming negligible at $|\vlsr| \geq 285$ \kms\ for both positive and negative velocities.    
We conclude that confusion has no more than a $10\%$ effect on our ability to detect  \hi\ clouds at $|\vlsr| \gtrsim 200 \ \kms$ within $5\dg$ of the Galactic plane.  In fact, confusion is actually less at $|\vlsr| \gtrsim 200 \ \kms$ than at velocities closer to zero, favoring detection of clouds with higher rather than lower $|\vlsr|$, exactly the opposite of the observed effect 
%\edt{Nice, I like this last sentence}.

\subsection{Sensitivity}
\label{sec:sensitivity}
The noise in a spectrum is a combination of contributions from instrumentation, thermal and non-thermal Galactic continuum  emission, and at some velocities, emission from the \hi\ line itself.   The Earth's atmosphere contributes a small bit as well.
Because the \hi\ clouds we study are detectable only at velocities lacking other bright \hi\ emission, the only important sources of noise for this discussion are instrumental, atmospheric,  and continuum emission.  
None of these components varies with frequency to an extent that would effect the detectability of \hi\ at different velocities.  
%The effect seen in Fig.~\ref{fig:AbsV_absb}.

\subsection{Change in cloud detectability with \texorpdfstring{$\mid\vlsr\mid$}{Lg}}
\label{sec:cloud_properties}
\begin{figure}
	\centering
	\includegraphics[width=0.48\textwidth]{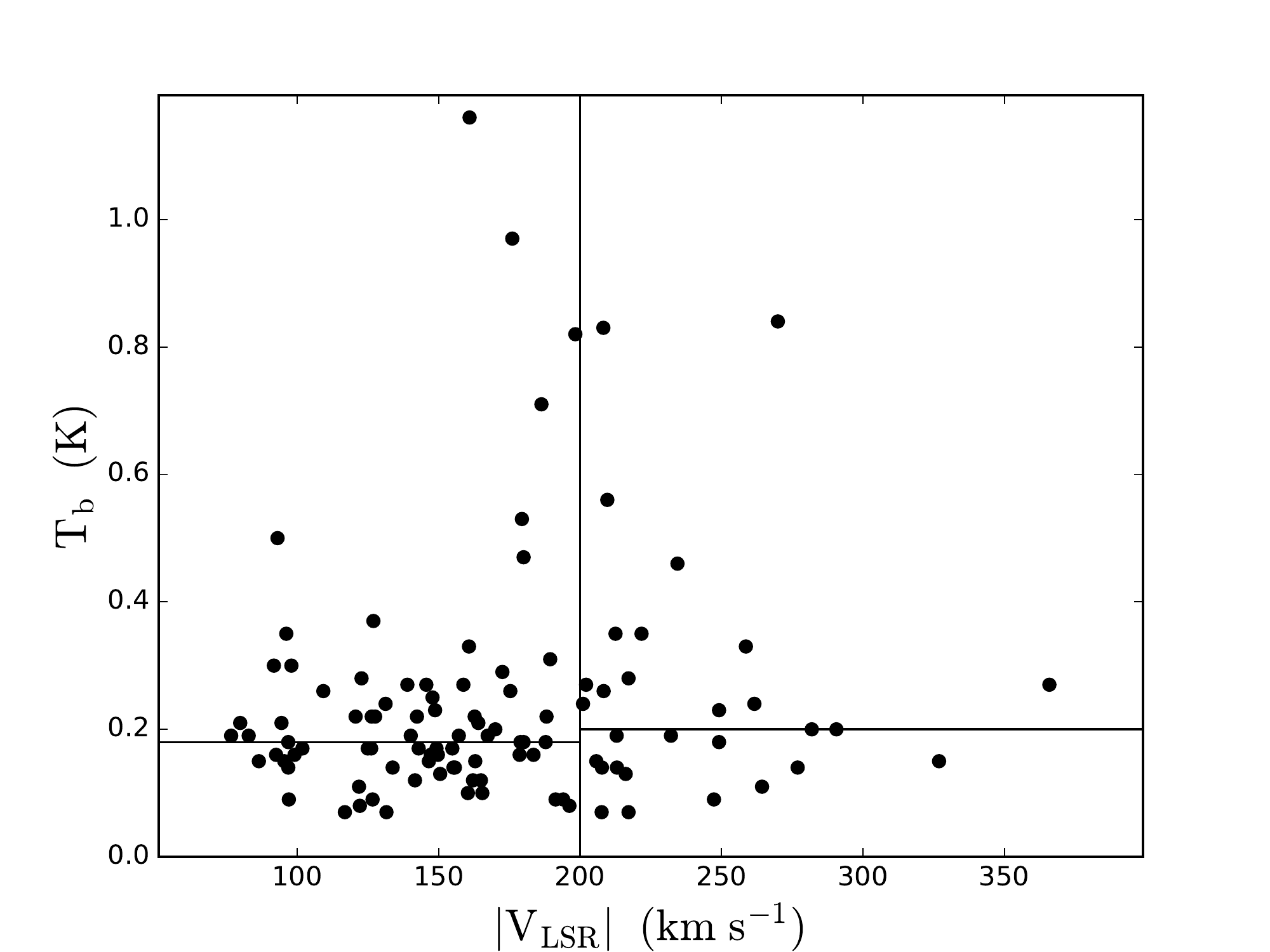}
	\caption{
		GBT cloud peak \hi\ line brightness temperature, $T_\mathrm{b}$, plotted against $|\vlsr|$ for clouds from the \citetalias{DiTeodoro+18} survey.
		Horizontal lines mark the median $T_\mathrm{b}$ for clouds with $|\vlsr| < 200$ \kms, and $|\vlsr| > 200$ \kms, respectively.  
		There is no significant difference between the line brightness, and hence the detectability, of clouds in these two velocity ranges.}
		\label{fig:Tb_vs_VLSR}
\end{figure}

The detectability of the clouds seen in the surveys depends mainly on their peak 21cm \hi\ brightness temperature, which is shown for the GBT sample in  \autoref{fig:Tb_vs_VLSR}  plotted against the $|\vlsr|$.  
Clouds to the right of the vertical line at 200 \kms\ have the large velocities that are not found  at $|b| \lesssim 4\dg$ (\autoref{fig:AbsV_absb}).  
The horizontal lines at 0.18 K and 0.20 K show the median \hi\ line brightness temperature  of clouds with  $|\vlsr|$ greater than and less than 200 \kms, respectively.   
There is no significant difference in the median peak line temperature, hence the detectability, of higher-velocity clouds compared to those at lower velocity.   
This conclusion also holds for the mean of the cloud line brightness temperature. 
We conclude that there is no evidence of a change in cloud detectability with $|\vlsr|$ that could produce the kinematic anomaly observed in \autoref{fig:AbsV_absb}.

\section{Acceleration of HI clouds?}
\label{sec:acceleration}

To explore the kinematics of the Fermi Bubble clouds we simulate a distribution of clouds within the confines of the \citet{MillerBregman16}  shell model  whose projected outlines are shown in \autoref{fig:HI_image}.  
In a right-handed cylindrical coordinate system, position is given by $(R, \theta, z)$ centered on the Galactic Center, where the Sun is at $R = R_0 \equiv 8.5$ kpc, $\theta_\odot = -180\dg$ moving on a purely circular orbit with $V_\odot = -220$ \kms.  The polar angle $\phi$ is measured from the plane in the direction of the $z$-axis  such that $\phi \equiv  \tan^{-1}(z/R)$.  The cone opening angle is then $\alpha = \pi - 2\phi_{\rm min}$.  The value of $\phi_{min}$ sets the boundary of the simulated cloud population close to the Galactic plane.  
Our goal is  to understand the kinematics of the clouds; we do not attempt to simulate their spatial distribution in this paper.
 
The observed \vlsr\ of a cloud moving with pure radial velocity \vout\ can then be written as:

\begin{align}
	\label{eq:vlsr}
	V_\mathrm{LSR} = \, &  V_\mathrm{out} \left[ \sin(\phi) \sin(b) -\cos(\phi)\cos(b)\cos(\ell+\theta)  \right] \nonumber \\
	& + V_\odot\sin(l)\cos(b) \quad . 
\end{align}

The values from \citetalias{DiTeodoro+18} of opening angle $\alpha = 140\arcdeg$ and a constant $\vout = 330$ \kms\  produce a $|\vlsr|$ distribution shown in \autoref{fig:Model_constVw}.  It does not reproduce the observed kinematic anomaly, as it has  clouds around $|\vlsr| = 300$ \kms\ at nearly all latitudes.  

The data provide some basic limits on  \vout\ and $\phi_{min}$. 
The outflow velocity must reach $\vout \gtrsim 300$ \kms\ at some locations to allow for values of $|\vlsr| \gtrsim 300$ \kms\ observed in the GBT data. 
Likewise, it is necessary that the angle $\phi_{min} \lesssim 20\dg$ or a number of the observed clouds will not lie within the boundaries of the cones.  
The Miller-Bregman shell model for the bubble volume actually implies $\phi_{min} = 0\dg$, i.e., an isotropic stream of clouds emanating from the Galactic Center.
In practice, because of confusion with other \hi\ emission at low $|b|$, the exact value of $\phi_{min}$ enters only weakly into the results,
but simulations with  $\phi_{min} < 15\dg$, i.e., a cone opening angle $\alpha > 150\dg$,
would predict the existence of clouds whose kinematics are not consistent with the kinematic anomaly for reasonable values of \vout.

 In the previous analyses, nearly identical values of  $\phi_{min}$ were found for the ATCA and GBT cloud samples (\citetalias{McClure-Griffiths+13}; \citetalias{DiTeodoro+18}) so this is unlikely be the source of the difference in their maximum $|\vlsr|$.  
We consider instead the case where the outflow velocity changes with distance from the Galactic Center, $d$.  For simplicity,  we consider models of the form 

\begin{equation}
    \label{eq:Vw_model}
	\vout (d) = 
    \begin{cases}
     V_0 + \left( V_\mathrm{max} - V_0\right)  \frac{d}{ d_\mathrm{acc}} & {\rm for \,\,\,} d < d_\mathrm{acc} \\ 
	 V_\mathrm{max} & {\rm for \,\,\,} d \geq d_\mathrm{acc}
    \end{cases}
\end{equation}

\noindent where $d$ is the distance from the Galactic Center and $d_\mathrm{acc}$ is the distance over which the outflow is accelerated from the minimum to the maximum value.  Note that for a purely radial outflow whose magnitude depends on $d$, the outflow velocity at any $|z|$ can vary considerably, especially at low $|z|$.

\begin{figure*}
	\centering
	\includegraphics[width=0.48\textwidth]{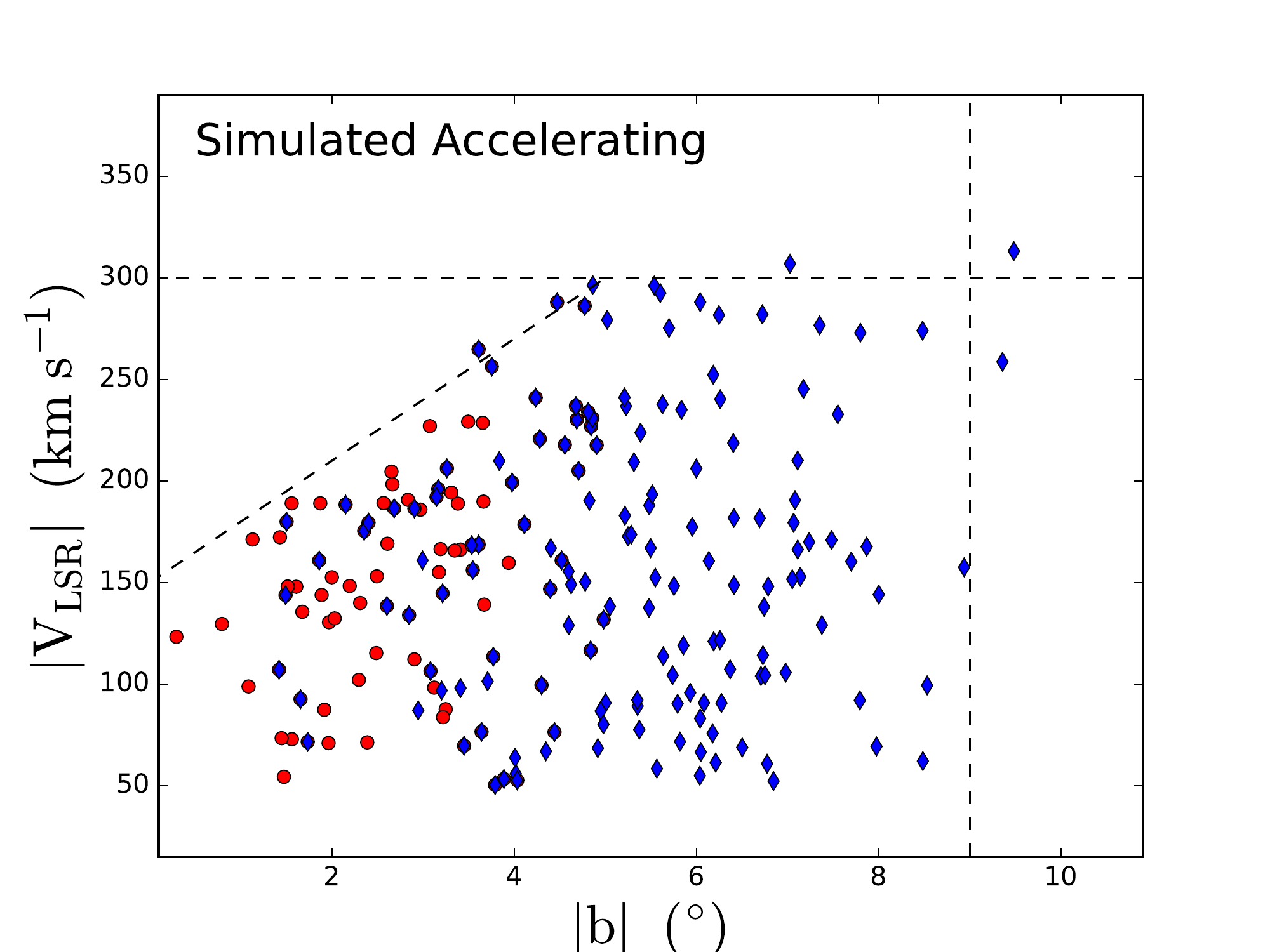}
	\hspace{0.02\textwidth}
	\includegraphics[width=0.48\textwidth]{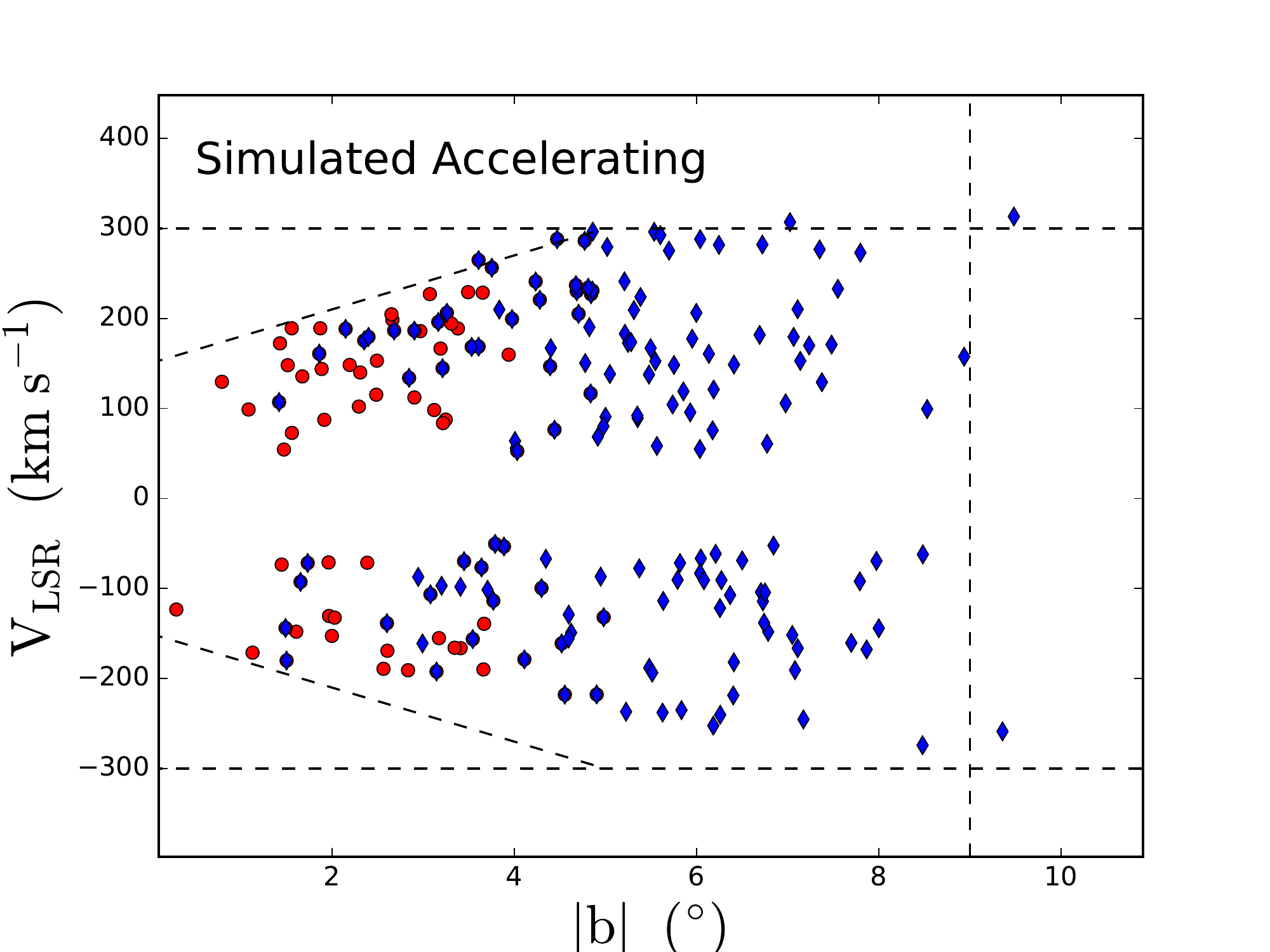}
	\caption{\emph{Left panel}: $|\vlsr|$ vs. $|b|$ for a simulated population of clouds within the Fermi Bubbles with an opening angle $\alpha = 145\dg$ and \vout\ that varies linearly from 175 \kms\ at $b = 0\dg$ to 330 \kms\ for clouds  3.5 kpc from the Galactic Center (model 3 of \autoref{tab:models}). 
	Red circles show clouds that appear in the regions covered only in the ATCA survey and blue diamonds show those that occur in regions covered by the GBT surveys.  
	The dashed fiducial lines are identical to those in \autoref{fig:AbsV_absb}, to which these figures should be compared.
	This simulation qualitatively matches the observations in that it reproduces the absence of clouds with  $|\vlsr| > 200$ \kms\  at $|b| \lesssim 4\arcdeg$. \emph{Right panel}: Same as left panel, but for positive and negative velocity clouds plotted separately. 
	}
	\label{fig:Model_linearVw}
\end{figure*}

\begin{deluxetable*}{ccccc}
	\tablecolumns{5}
	\tablecaption{Simulations that give acceptable agreement with the observations.   The parameter $\alpha$ is the cone opening angle; other parameters  are defined in Eq.~1 and Eq.~2. 
	\label{tab:models}}
	\tablehead{\colhead{Model} &
		\colhead{$V_0$} & 
		\colhead{$V_{max}$}  & 
		\colhead{$d_{acc}$} & 
		\colhead{$\alpha$} \\
	    \colhead{} & 	\colhead{(\kms)} & \colhead{(\kms)} & \colhead{(\kms\ kpc$^{-1}$)} & \colhead{($\arcdeg$)}  \\
		\colhead{(1)} &  \colhead{(2)} & \colhead{(3)} & \colhead{(4)}  & \colhead{(5)}
	}
 \startdata
\hline\hline\noalign{\vspace{5pt}}	
\noalign{\smallskip}
1 & 165 & 330 & 2.5 & 140\\
2 & 165 & 300 & 3.5 & 150\\
3 & 175 & 330 & 3.5 & 145\\
4 & 200 & 330 & 2.5 & 140\\
 \noalign{\vspace{5pt}}
 \enddata
\end{deluxetable*}

A set of simulations was run with varying parameters  and compared with properties of the 196 known \hi\ clouds.  The simulations were evaluated on three criteria: 1) The parameters had to produce a distribution of $\vlsr$ with $|b|$ that reproduced the kinematic anomaly.  2)  Most of the  clouds in the ATCA and GBT surveys had to be consistent with  the ($\ell,b,\vlsr$) limits of the simulation.  
3) At the $(\ell,b$) of each observed cloud, the simulation had to produce a $\vlsr$ within a few \kms\ of the observed \vlsr\ somewhere within the adopted Fermi Bubble boundaries. 
There was no attempt to match the distribution of the numbers of observed clouds in  longitude or latitude, only their kinematics.

A selection of parameter combinations that produces acceptably good distributions is given in \autoref{tab:models}.
These models can account for  the kinematics of $\gtrsim95\%$ of the observed \hi\ clouds as well as the components in the 
UV absorption spectra.  
\autoref{fig:Model_linearVw} shows detailed results for one set of parameters for comparison with \autoref{fig:AbsV_absb}.  
We conclude that the  simplest explanation for the kinematic anomaly of Fermi Bubble \hi\ clouds observed at low $|b|$ is an increase in \vout\ with distance from the Galactic Center.

	We have explored simulations where the clouds arise from a region several hundred pc in size around the Galactic Center and not from a single point, but those still require acceleration to produce the kinematic anomaly.  It is important to realize that the detected \hi\ clouds extend to the limits of the areas covered by the ATCA and GBT surveys so the full extent of the cloud population in longitude and latitude is not known.   The GASS survey does cover a much larger area, but we have found that it is not very good at detecting clouds such as those found by the GBT.  This is most likely because its poorer angular resolution causes significant beam dilution of the 21cm line; it also has  lower sensitivity.  The current \hi\ data thus do not place significant constraints on the size of the region that originates the clouds.    	A similar situation applies to the cloud acceleration.  For simplicity, we have adopted a linear acceleration term  because the data provide poor constraints on more complex (and more realistic) acceleration functions. More extensive observations are required. 

\section{Discussion and Conclusions}
\label{sec:discussion}

Although the simulations we use here are quite crude, they clearly imply that the 
 clouds observed in the lower regions of the Fermi Bubbles 
 must have been accelerated from $\lesssim 200$ \kms\ to  $\approx 330$ \kms\ in their  first few kpc of travel away from the Galactic Center. 
Using the kinematic model of Eq.\ref{eq:Vw_model} we can estimate a  location for every \hi\ cloud in the ATCA and GBT catalogs.   
Virtually  all lie at $|z| < 1.5$ kpc at a distance $d < 3$ kpc from the Galactic Center and have presumably been 
entrained in a hot wind that has a much higher expansion velocity.  
Components detected in UV spectra at $b = -6\fdg6$ and $b = +11\fdg2$ over a wide range of ionization stage \citep{Fox+15,Savage+17}  have kinematics consistent with those of the \hi\ clouds.

Observations, theoretical studies, and hydrodynamical simulations suggest that a hot wind can accelerate compact, cold gas clouds such as we observe in the 21cm line \citep[e.g.,][]{Castor+1975,Veilleux+05,Cooper+08}.
In the so-called ``entrainment'' scenario, cold gas is hit by the hot flow and dragged until it is either shredded or approaches the velocity of the hot gas. 
The details of such a process are still unclear and very debated. 
Many high-resolution simulations of cold gas clouds in hot flows have shown that entrainment is a highly destructive process, the cold gas being ripped and ionized on time-scales much shorter than needed to reach the observed velocities of hundreds of \kms\  \citep[e.g.][]{Scannapieco&Bruggen15,Schneider+17,Sparre+19}.
In fact, when the density of the cold gas is much higher than the density of the hot gas, the cloud-crushing time $t_\mathrm{cc}$, which governs the destruction timescale, is always much lower than the acceleration time $t_\mathrm{acc}$ because of the drag force, i.e.\ $t_\mathrm{cc}\ll t_\mathrm{acc}$.
However, some studies suggest that a cloud lifetime can be prolonged through radiative cooling \citep{Gronke+18}, thermal conduction \citep{Bruggen&Scannapieco16} and magnetic fields \citep{McCourt+15}.

The kinematic model we use to explore $\vlsr$ can also be used to estimate cloud lifetimes, though  
 the form of the acceleration is not well constrained by our data.   Whereas we assume a linear acceleration, in an 
entrainment scenario acceleration is expected to be more violent \citep[e.g.,][]{Zhang+17}.
Nonetheless, our values for $V_0$ and $V_{\rm max}$ give estimated cloud lifetimes slightly longer than those calculated with a constant outflow velocity  by 
\citetalias{DiTeodoro+18} and range from 4 Myr to 10 Myr, with a median of 6.5 Myr. 
These longer lifetimes add an additional challenge to the entrainment scenario.

 While the ~200 \hi\ clouds that we study here appear to be part of a single population whose kinematics can be described quite simply, there is certainly the possibility that the Galactic nucleus undergoes periodic outbursts \citep[e.g.,][]{Veilleux+05,Bland-Hawthorn+19} leaving other anomalous-velocity gas as its mark on the interstellar medium.  This might explain the anomalous high-velocity absorption observed toward the Galactic center at very high latitudes \citep{Bordoloi+17,Karim+18} whose kinematics is not consistent with the outflow described here.

As is evident in \autoref{fig:HI_image}, only a fraction of the volume of the Fermi Bubbles has been searched for \hi\ clouds to the sensitivity levels needed to detect the objects described here and to follow their evolution in space.  We are continuing our 21cm \hi\ observational program with the GBT to further refine the properties of this population.
Additional measurements of UV absorption spectra through the Fermi Bubbles  in the range $10\dg \leq |b| \leq 30\dg$ would supply critical information.

\acknowledgments{
Observations were made with the Green Bank Telescope under proposal 19A\_337. The Green Bank Observatory is a facility of the National Science Foundation, operated under a cooperative agreement by Associated Universities, Inc.
We acknowledge the support of the Australian Research Council (ARC) through grant DP160100723. N.M.-G.\ acknowledges the support of the ARC through Future Fellowship FT150100024.
}

\facilities{ATCA, GBT, Parkes} 
\software{$^\mathrm{3D}$\textsc{Barolo} \citep{DiTeodoro+15}, \textsc{Gbtgridder}, \textsc{GBTIDL}, \textsc{Miriad} \citep*{Sault+95}. }
\vspace*{2cm}

\bibliography{acceleration}

\begin{thebibliography}{}
\expandafter\ifx\csname natexlab\endcsname\relax\def\natexlab#1{#1}\fi
\providecommand{\url}[1]{\href{#1}{#1}}
\providecommand{\dodoi}[1]{doi:~\href{http://doi.org/#1}{\nolinkurl{#1}}}
\providecommand{\doeprint}[1]{\href{http://ascl.net/#1}{\nolinkurl{http://ascl.net/#1}}}
\providecommand{\doarXiv}[1]{\href{https://arxiv.org/abs/#1}{\nolinkurl{https://arxiv.org/abs/#1}}}

\bibitem[{{Binney} {et~al.}(1991){Binney}, {Gerhard}, {Stark}, {Bally}, \&
  {Uchida}}]{Binney+91}
{Binney}, J., {Gerhard}, O.~E., {Stark}, A.~A., {Bally}, J., \& {Uchida}, K.~I.
  1991, \mnras, 252, 210, \dodoi{10.1093/mnras/252.2.210}

\bibitem[{{Bland-Hawthorn} \& {Cohen}(2003)}]{Bland-Hawthorn_Cohen03}
{Bland-Hawthorn}, J., \& {Cohen}, M. 2003, \apj, 582, 246,
  \dodoi{10.1086/344573}

\bibitem[{{Bland-Hawthorn} {et~al.}(2019){Bland-Hawthorn}, {Maloney},
  {Sutherland}, {Groves}, {Guglielmo}, {Hao Li}, {Curzons}, {Cecil}, \&
  {Fox}}]{Bland-Hawthorn+19}
{Bland-Hawthorn}, J., {Maloney}, P., {Sutherland}, R., {et~al.} 2019, arXiv
  e-prints, arXiv:1910.02225.
\newblock \doarXiv{1910.02225}

\bibitem[{{Boothroyd} {et~al.}(2011){Boothroyd}, {Blagrave}, {Lockman},
  {Martin}, {Pinheiro Gon{\c c}alves}, \& {Srikanth}}]{Boothroyd+11}
{Boothroyd}, A.~I., {Blagrave}, K., {Lockman}, F.~J., {et~al.} 2011, \aap, 536,
  A81, \dodoi{10.1051/0004-6361/201117656}

\bibitem[{{Bordoloi} {et~al.}(2017){Bordoloi}, {Fox}, {Lockman}, {Wakker},
  {Jenkins}, {Savage}, {Hernandez}, {Tumlinson}, {Bland-Hawthorn}, \&
  {Kim}}]{Bordoloi+17}
{Bordoloi}, R., {Fox}, A.~J., {Lockman}, F.~J., {et~al.} 2017, \apj, 834, 191,
  \dodoi{10.3847/1538-4357/834/2/191}

\bibitem[{{Br{\"u}ggen} \& {Scannapieco}(2016)}]{Bruggen&Scannapieco16}
{Br{\"u}ggen}, M., \& {Scannapieco}, E. 2016, \apj, 822, 31,
  \dodoi{10.3847/0004-637X/822/1/31}

\bibitem[{{Burton} \& {Liszt}(1978)}]{Burton&Liszt78}
{Burton}, W.~B., \& {Liszt}, H.~S. 1978, \apj, 225, 815, \dodoi{10.1086/156547}

\bibitem[{{Carretti} {et~al.}(2013){Carretti}, {Crocker}, {Staveley-Smith},
  {Haverkorn}, {Purcell}, {Gaensler}, {Bernardi}, {Kesteven}, \&
  {Poppi}}]{Carretti+13}
{Carretti}, E., {Crocker}, R.~M., {Staveley-Smith}, L., {et~al.} 2013, \nat,
  493, 66, \dodoi{10.1038/nature11734}

\bibitem[{{Castor} {et~al.}(1975){Castor}, {McCray}, \& {Weaver}}]{Castor+1975}
{Castor}, J., {McCray}, R., \& {Weaver}, R. 1975, \apjl, 200, L107,
  \dodoi{10.1086/181908}

\bibitem[{{Cooper} {et~al.}(2008){Cooper}, {Bicknell}, {Sutherland}, \&
  {Bland-Hawthorn}}]{Cooper+08}
{Cooper}, J.~L., {Bicknell}, G.~V., {Sutherland}, R.~S., \& {Bland-Hawthorn},
  J. 2008, \apj, 674, 157, \dodoi{10.1086/524918}

\bibitem[{{Di Teodoro} \& {Fraternali}(2015)}]{DiTeodoro+15}
{Di Teodoro}, E.~M., \& {Fraternali}, F. 2015, \mnras, 451, 3021,
  \dodoi{10.1093/mnras/stv1213}

\bibitem[{{Di Teodoro} {et~al.}(2018){Di Teodoro}, {McClure-Griffiths},
  {Lockman}, {Denbo}, {Endsley}, {Ford}, \& {Harrington}}]{DiTeodoro+18}
{Di Teodoro}, E.~M., {McClure-Griffiths}, N.~M., {Lockman}, F.~J., {et~al.}
  2018, \apj, 855, 33, \dodoi{10.3847/1538-4357/aaad6a}

\bibitem[{{Dobler} \& {Finkbeiner}(2008)}]{Dobler&Finkbeiner08}
{Dobler}, G., \& {Finkbeiner}, D.~P. 2008, \apj, 680, 1222,
  \dodoi{10.1086/587862}

\bibitem[{{Fox} {et~al.}(2015){Fox}, {Bordoloi}, {Sav age}, {Lockman},
  {Jenkins}, {Wakker}, {Bland-Hawthorn}, {Hernandez}, {Kim}, {Benjamin},
  {Bowen}, \& {Tumlinson}}]{Fox+15}
{Fox}, A.~J., {Bordoloi}, R., {Sav age}, B.~D., {et~al.} 2015, \apjl, 799, L7,
  \dodoi{10.1088/2041-8205/799/1/L7}

\bibitem[{{Gronke} \& {Oh}(2018)}]{Gronke+18}
{Gronke}, M., \& {Oh}, S.~P. 2018, \mnras, 480, L111,
  \dodoi{10.1093/mnrasl/sly131}

\bibitem[{{Kalberla} {et~al.}(2010){Kalberla}, {McClure-Griffiths}, {Pisano},
  {Calabretta}, {Ford}, {Lockman}, {Staveley-Smith}, {Kerp}, {Winkel},
  {Murphy}, \& {Newton-McGee}}]{Kalberla+10}
{Kalberla}, P.~M.~W., {McClure-Griffiths}, N.~M., {Pisano}, D.~J., {et~al.}
  2010, \aap, 521, A17, \dodoi{10.1051/0004-6361/200913979}

\bibitem[{{Karim} {et~al.}(2018){Karim}, {Fox}, {Jenkins}, {Bordoloi},
  {Wakker}, {Savage}, {Lockman}, {Crawford}, {Jorgenson}, \&
  {Bland-Hawthorn}}]{Karim+18}
{Karim}, M.~T., {Fox}, A.~J., {Jenkins}, E.~B., {et~al.} 2018, \apj, 860, 98,
  \dodoi{10.3847/1538-4357/aac167}

\bibitem[{{Kataoka} {et~al.}(2013){Kataoka}, {Tahara}, {Totani}, {Sofue},
  {Stawarz}, {Takahashi}, {Takeuchi}, {Tsunemi}, {Kimura}, {Takei}, {Cheung},
  {Inoue}, \& {Nakamori}}]{Kataoka+13}
{Kataoka}, J., {Tahara}, M., {Totani}, T., {et~al.} 2013, \apj, 779, 57,
  \dodoi{10.1088/0004-637X/779/1/57}

\bibitem[{{Keeney} {et~al.}(2006){Keeney}, {Danforth}, {Stocke}, {Penton},
  {Shull}, \& {Sembach}}]{Keeney+06}
{Keeney}, B.~A., {Danforth}, C.~W., {Stocke}, J.~T., {et~al.} 2006, \apj, 646,
  951, \dodoi{10.1086/505128}

\bibitem[{{Lockman}(1984)}]{Lockman84}
{Lockman}, F.~J. 1984, \apj, 283, 90, \dodoi{10.1086/162277}

\bibitem[{{Lockman} \& {McClure-Griffiths}(2016)}]{Lockman_McClureGriffiths16}
{Lockman}, F.~J., \& {McClure-Griffiths}, N.~M. 2016, \apj, 826, 215,
  \dodoi{10.3847/0004-637X/826/2/215}

\bibitem[{{McClure-Griffiths} {et~al.}(2013){McClure-Griffiths}, {Green},
  {Hill}, {Lockman}, {Dickey}, {Gaensler}, \& {Green}}]{McClure-Griffiths+13}
{McClure-Griffiths}, N.~M., {Green}, J.~A., {Hill}, A.~S., {et~al.} 2013,
  \apjl, 770, L4, \dodoi{10.1088/2041-8205/770/1/L4}

\bibitem[{{McClure-Griffiths} {et~al.}(2009){McClure-Griffiths}, {Pisano},
  {Calabretta}, {Ford}, {Lockman}, {Staveley-Smith}, {Kalberla}, {Bailin},
  {Dedes}, {Janowiecki}, {Gibson}, {Murphy}, {Nakanishi}, \&
  {Newton-McGee}}]{McClure-Griffiths+09}
{McClure-Griffiths}, N.~M., {Pisano}, D.~J., {Calabretta}, M.~R., {et~al.}
  2009, \apjs, 181, 398, \dodoi{10.1088/0067-0049/181/2/398}

\bibitem[{{McCourt} {et~al.}(2015){McCourt}, {O'Leary}, {Madigan}, \&
  {Quataert}}]{McCourt+15}
{McCourt}, M., {O'Leary}, R.~M., {Madigan}, A.-M., \& {Quataert}, E. 2015,
  \mnras, 449, 2, \dodoi{10.1093/mnras/stv355}

\bibitem[{{Miller} \& {Bregman}(2016)}]{MillerBregman16}
{Miller}, M.~J., \& {Bregman}, J.~N. 2016, \apj, 829, 9,
  \dodoi{10.3847/0004-637X/829/1/9}

\bibitem[{{Oort}(1977)}]{Oort77}
{Oort}, J.~H. 1977, \araa, 15, 295, \dodoi{10.1146/annurev.aa.15.090177.001455}

\bibitem[{{Sault} {et~al.}(1995){Sault}, {Teuben}, \& {Wright}}]{Sault+95}
{Sault}, R.~J., {Teuben}, P.~J., \& {Wright}, M.~C.~H. 1995, in Astronomical
  Society of the Pacific Conference Series, Vol.~77, Astronomical Data Analysis
  Software and Systems IV, ed. R.~A. {Shaw}, H.~E. {Payne}, \& J.~J.~E.
  {Hayes}, 433

\bibitem[{{Savage} {et~al.}(2017){Savage}, {Kim}, {Fox}, {Massa}, {Bordoloi},
  {Jenkins}, {Lehner}, {Bland-Hawthorn}, {Lockman}, {Hernandez}, \&
  {Wakker}}]{Savage+17}
{Savage}, B.~D., {Kim}, T.-S., {Fox}, A.~J., {et~al.} 2017, \apjs, 232, 25,
  \dodoi{10.3847/1538-4365/aa8f4c}

\bibitem[{{Scannapieco} \& {Br{\"u}ggen}(2015)}]{Scannapieco&Bruggen15}
{Scannapieco}, E., \& {Br{\"u}ggen}, M. 2015, \apj, 805, 158,
  \dodoi{10.1088/0004-637X/805/2/158}

\bibitem[{{Schneider} \& {Robertson}(2017)}]{Schneider+17}
{Schneider}, E.~E., \& {Robertson}, B.~E. 2017, \apj, 834, 144,
  \dodoi{10.3847/1538-4357/834/2/144}

\bibitem[{{Sparre} {et~al.}(2019){Sparre}, {Pfrommer}, \&
  {Vogelsberger}}]{Sparre+19}
{Sparre}, M., {Pfrommer}, C., \& {Vogelsberger}, M. 2019, \mnras, 482, 5401,
  \dodoi{10.1093/mnras/sty3063}

\bibitem[{{Su} {et~al.}(2010){Su}, {Slatyer}, \& {Finkbeiner}}]{Su+10}
{Su}, M., {Slatyer}, T.~R., \& {Finkbeiner}, D.~P. 2010, \apj, 724, 1044,
  \dodoi{10.1088/0004-637X/724/2/1044}

\bibitem[{{Veilleux} {et~al.}(2005){Veilleux}, {Cecil}, \&
  {Bland-Hawthorn}}]{Veilleux+05}
{Veilleux}, S., {Cecil}, G., \& {Bland-Hawthorn}, J. 2005, \araa, 43, 769,
  \dodoi{10.1146/annurev.astro.43.072103.150610}

\bibitem[{{Zhang} {et~al.}(2017){Zhang}, {Thompson}, {Quataert}, \&
  {Murray}}]{Zhang+17}
{Zhang}, D., {Thompson}, T.~A., {Quataert}, E., \& {Murray}, N. 2017, \mnras,
  468, 4801, \dodoi{10.1093/mnras/stx822}

\end{thebibliography}
\bibliographystyle{aasjournal}

\appendix
\section{GBT Observations and Data Reduction}
\label{appendix:data}

Observations of the 21 cm line were made with the Robert C. Byrd Green Bank Telescope (GBT) of the Green Bank Observatory that has an angular resolution of $9\farcm1$ at 1.4 GHz.
The GBT L-band receiver used for these measurements has a system temperature on cold sky of 18 K, although at low latitudes near the Galactic Center nonthermal continuum emission increases the system temperature significantly in some directions.
Spectra were taken using in-band frequency-switching which gave a clean spectral response over $\pm 656$ \kms\ at a channel spacing of 0.15 \kms. 
Maps were made on-the-fly scanning in Galactic longitude at a fixed latitude.  The region between $0\dg \leq \ell \leq 6\dg$, $+1\dg \leq b \leq +3\dg$ was covered in 3 patches, each $2\dg \times 2\dg$. Each patch was observed several times to reduce the noise level for a total integration time of 46 hours over the 12 square degree field.

Spectra were calibrated, smoothed to 1.06  \kms\ velocity resolution, and corrected for stray radiation using the procedure described by \citet{Boothroyd+11}, and a third degree polynomial was fit to emission-free regions of each spectrum.

All of the clouds first detected over this region in the ATCA survey were present in the new GBT data. In addition, the cloud-finding algorithm described in \citetalias{DiTeodoro+18} isolated four new clouds, whose properties are listed in \autoref{tab:newGBTdata}. 
Although the newly-detected clouds all have negative velocities this does not seem to be the result of any bias.  When all clouds in this area are included, nine have negative and 12 have positive velocities.

\end{document}